\documentclass[10pt,aps,prb,twocolumn,preprintnumbers,amsmath,amssymb,floatfix,citeautoscript]{revtex4-1}
\usepackage[latin1]{inputenc}
\usepackage[T1]{fontenc}
\usepackage[english]{babel}
\usepackage[pdftex]{graphicx}
\usepackage{amsmath}
\usepackage{times}
\usepackage{psfrag}
\usepackage{csquotes}
\usepackage[usenames,dvipsnames,svgnames,table]{xcolor}
\usepackage[colorlinks,plainpages=false,linkcolor=blue,urlcolor=blue,citecolor=blue,pdfpagemode=UseNone]{hyperref}

\renewcommand{\AA}{\text{\r{A}}}

\newcommand\Int[2]{\int_{#1} \, \text{d}#2 \,}

\newcommand\Exp{\text{e}}
\newcommand\Multp{\cdot}
\newcommand\Vek[1]{\vec{#1}}

\begin{document}

\title
{
Spincaloric properties of epitaxial Co$_2$MnSi/MgO/Co$_2$MnSi magnetic tunnel junctions
}

\author{Benjamin Geisler}
\email{benjamin.geisler@uni-due.de}
\author{Peter Kratzer}

\affiliation{
Fakult\"at f\"ur Physik and Center for Nanointegration (CENIDE), Universit\"at Duisburg-Essen, 47048 Duisburg, Germany
}

\date{\today}

\begin{abstract}
The electronic transport and spincaloric properties of
epitaxial magnetic tunnel junctions with half-metallic Co$_2$MnSi Heusler electrodes, MgO tunneling barriers, and different interface terminations are investigated by using first-principles calculations.
A new approach to spincaloric properties is presented that circumvents the linear response approximation inherent in the Seebeck coefficient and compared to the method of Sivan and Imry.
This approach supports two different temperatures in the two electrodes
and provides the exact current and/or voltage response of the system.
Moreover, it accounts for temperature-dependent chemical potentials in the electrodes and finite-bias effects.
We find that especially the former are important for obtaining qualitatively correct results,
even if the variations of the chemical potentials are small.
It is shown how the spincaloric properties can be tailored by the choice of the growth conditions.
We find a large effective and spin-dependent Seebeck coefficient of $-65$~$\mu$V/K at room temperature for the purely Co-terminated interface.
We suggest to use
such interfaces in thermally operated magnetoresistive random access memory modules,
which exploit the magneto-Seebeck effect, to maximize the thermally induced readout voltage.
\end{abstract}

\maketitle

\section{Introduction}

Magnetic tunnel junctions (MTJs) with ferromagnetic, half-metallic electrodes
are interesting spintronics~\cite{Prinz:98, Wolf:01, SpintronicsFabian:04} devices due to their high tunnel magnetoresistance (TMR) ratio;
if a voltage is applied to such a device, the resulting current depends strongly on the relative magnetization of the electrodes and ideally vanishes for the antiparallel configuration.
Thus, MTJs can store information and are, for instance, building blocks of magnetoresistive random access memory (MRAM).

\begin{figure}[t]
	\centering
	\includegraphics[]{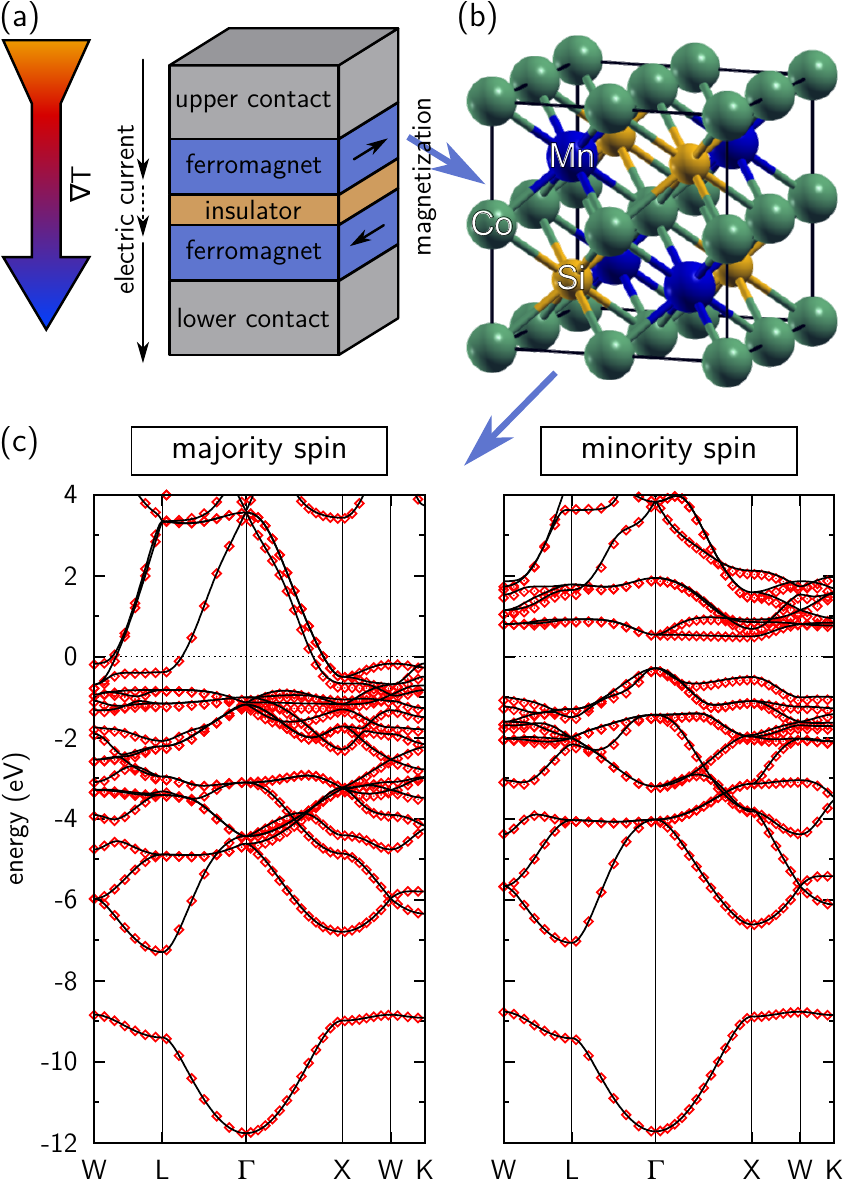}
	\caption{(Color online) (a)~Schematic illustration of a MTJ. Either an electric field or a thermal gradient can be applied to the device. (b)~Atomic structure of bulk Co$_2$MnSi and (c)~its electronic band structure for the two different spin channels. The band gap in the minority spin channel at the Fermi energy (zero energy) is clearly visible. The red diamonds are our all-electron LAPW results for comparison, underlining the accuracy of the pseudopotential approach (black lines).}
	\label{fig:Heusler-Bands}
\end{figure}

This stored information can also be read out by application of a thermal gradient instead of an electric field.
The Seebeck voltage, which arises in a MTJ due to a thermal gradient between the two electrodes [cf.~Fig.~\ref{fig:Heusler-Bands}(a)], can be used to detect the state of the electrode magnetization~\cite{Heiliger:11}.
This magneto-Seebeck effect can be expected to be very large in the case of half-metallic electrodes.
In contrast to conventional MRAM modules~\cite{Akerman:05}, no charge current flows in the readout process.
Hence, aging effects in the devices due to electromigration can be reduced.

There has been quite some interest in epitaxial Co$_2$MnSi/MgO$(001)$/Co$_2$MnSi MTJs in the past, both experimental and theoretical.
Ishikawa \textit{et al.}\ recently reported a TMR ratio of $705\,\%$ at $4.2$~K and $182\,\%$ at room temperature~\cite{Ishikawa:09},
while Liu \textit{et al.}\ achieved almost $2000\,\%$ at $4.2$~K and up to $350\,\%$ at room temperature (with an additional CoFe buffer layer as substrate for the lower electrode~\cite{Liu:12}).
H\"ulsen \textit{et al.}\ reported on the electronic structure of different Co$_2$MnSi/MgO$(001)$ interfaces~\cite{HuelsenPRL:09}.

In the present paper we will approach this system from a different angle.
\textit{Ab initio} electronic transport calculations are employed to investigate thermoelectric and/or spincaloric properties in dependence on the interface atomic structure.
In order to obtain enough data to reliably deduce these quantities,
previous transport calculations by Miura \textit{et al.}~\cite{Miura:08, Miura:09}
that are conceptually similar to our work
had to be extended considerably.
The interface atomic structure can be influenced by the growth conditions~\cite{HuelsenPRL:09},
which provides the opportunity to tailor and optimize the spincaloric properties in real MTJ devices.
We compare results calculated by using the conventionally employed, approximate method of Sivan and Imry~\cite{SI86}
with results obtained from the Landauer-B\"uttiker equation~\cite{Buettiker:85}.
The latter procedure, which we introduce in this paper, circumvents the linear response approximation inherent in the Seebeck coefficient
and directly provides the response of the system (current or voltage) to arbitrary electrode temperatures.
Moreover, thermal variations of the chemical potentials in the electrodes and finite-bias effects can be readily included in this method.
We find that the former, albeit being small, lead to considerable quantitative \textit{and} qualitative differences in the thermally induced current and voltage from expectations based solely on the conventional Seebeck coefficient.
Finally, we present the concept of thermally operated MRAM modules,
which exploit the magneto-Seebeck effect,
and provide an estimate of the expected voltages in these devices under realistic conditions.

\section{Numerical details}

The electronic structure and transport calculations 
have been performed within the framework of spin-polarized 
density functional theory~\cite{KoSh65} (DFT)
employing the plane-wave pseudopotential method 
as implemented in the {\sc Quan\-tum\-Es\-pres\-so} code~\cite{PWSCF},
with the PBE generalized gradient approximation
parametrization of the exchange-correlation functional~\cite{PeBu96}.
Wave functions and density have been expanded into plane waves
up to cutoff energies of $35$ and $350$~Ry, respectively.
The neighborhood of atom centers has been approximated by
self-created ultrasoft pseudopotentials (USPPs~\cite{Vanderbilt:1990}),
treating the atomic
Co $3d$, $4s$, $4p$,
Mn $3p$, $3d$, $4s$, $4p$,
Si $3s$, $3p$,
Mg $2p$, $3s$, $3p$,
and O $2s$, $2p$
subshells as valence states~\cite{Geisler:12, Geisler:13}.
For Co, Mn, and Si, a nonlinear core correction~\cite{LoFr82} has been included.
During the pseudopotential creation process a 
scalar-relativistic approximation has been applied to
the electron motion. 
A Methfessel-Paxton smearing~\cite{MePa89} of $10$~mRy
has been used during the Brillouin zone (BZ) sampling,
which has been performed
with a $16 \times 16 \times 2$ Monkhorst-Pack $k$-point grid~\cite{MoPa76}
for the Heusler/MgO/Heusler supercells
and a $16 \times 16 \times 12$ $k$-point grid for the Heusler electrodes.
All grids have been chosen in such a way
that they do not include the $\Gamma$ point
and deliver accurately converged Fermi energies and
potentials.

All internal atomic positions have been accurately optimized by using
Hellmann-Feynman forces to reduce the force components below
$1$~mRy/bohr and the energy changes below $0.1$~mRy.
Moreover, the length of every considered Heusler/MgO/Heusler supercell
has been optimized in order to determine
the ideal, energy-minimizing Heusler-MgO spacing for each interface termination.

For the transport properties, we have considered an open quantum
system consisting of
(i)~a scattering region comprising the MgO barrier material
and a chosen interface to the Heusler electrodes, and 
(ii)~the left and right semi-infinite Heusler electrodes (leads).
From the accurately converged DFT potentials of the leads and of the
scattering region,
transport coefficients have been calculated separately for
both spin channels by using a method following
Refs.~\onlinecite{SmogunovTosatti:04}
and~\onlinecite{ChoiIhm:99}.
In order to sample the two-dimensional (2D) BZ
(perpendicular to the direction of the tunneling current)
on a reasonable computational time scale, we have 
massively parallelized the method. 
Sufficient convergence of the energy- and spin-resolved
transmission,
\begin{equation}
	\label{eq:Heusler-TofE}
	{\cal T}_\sigma(E) = \frac{1}{A_{\text{BZ}}} \int \text{d}^2 k_{\perp}
	\,
	{\cal T}_\sigma(E, \Vek{k}_{\perp})
\text{,}
\end{equation}
%
with respect to the
2D $k_{\perp}$-point grid has been found to be attained with
a $401 \times 401$ regular mesh.
Here, $A_{\text{BZ}}$ is the area of the 2D BZ.
The regular energy mesh on which ${\cal T}_\sigma(E)$ has been explicitly calculated has a spacing of $25$~meV.
Subsequently, the transmission has been interpolated on a refined energy mesh with a $1.36$~meV ($0.1$~mRy) spacing.
As we will see in the following,
the transmission ${\cal T}_\sigma(E)$ is the central quantity
in all subsequent considerations.

The focus of this paper lies on the electronic transport through the MTJs for a parallel magnetization of the ferromagnetic electrodes.
We neglect potential contributions of Co$_2$MnSi nonquasiparticle states near the conduction band of the minority spin channel~\cite{Chioncel-NQP:08}
as well as finite-temperature inelastic processes, e.g., due to phonons or (interface) magnons~\cite{Miura:11, Akai:2012, Mavropoulos:06}.
Especially the latter are suspected to induce a small, finite transmission of minority spin electrons despite the half-metallic band gap, thus reducing the TMR ratio or the spin-dependent Seebeck effect at larger temperatures.
Since only light elements are involved, we neglect the influence of
spin-orbit interaction, which can (i)~lead to a very small finite density
of states within the half-metallic band gap~\cite{Heusler-SOC-Mavropoulos:04} and (ii)~give rise to
a small spin-flip scattering.
It is possible to include the effect of spin disorder on the spincaloric phenomena,
as it has been done, for instance, for nanostructured Co systems~\cite{Mavropoulos:14}
or (Cr,Zn)Te half-metallic nanostructures~\cite{Mavropoulos:15}.
Note, however, that the Curie temperature of Co$_2$MnSi is three times as high as for CrTe ($334$~K),
which is why such effects are expected to be far more important in the latter case than in the former.
Finally, we make the assumption
that the process of electron tunneling through the insulating barriers occurs at a lower rate than energy dissipation and thermalization processes in the electrodes (reservoirs),
so that we have well-defined temperatures, chemical potentials, and Fermi-like distribution functions in the electrodes at all times.

\begin{figure*}[t]
	\centering
	\includegraphics[]{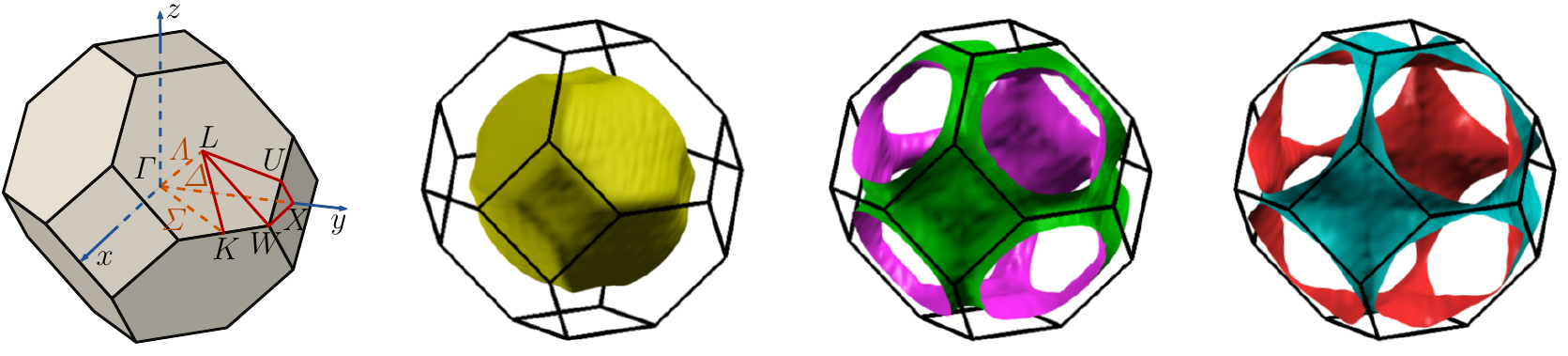}
	\caption{(Color online) Majority spin Fermi surface sheets of bulk Co$_2$MnSi, shown in the first BZ of the fcc lattice. They have been derived from the PBE electronic structure and correspond to the three different bands that cross the Fermi energy [cf.~Fig.~\ref{fig:Heusler-Bands}(c)].}
	\label{fig:Heusler-FermiSurfaces}
\end{figure*}

\section{Atomic and magnetic structure}

\subsection{The bulk Heusler material}

The ferromagnetic ($T_{\text{C}} = 985$~K~\cite{Webster:71}) ternary Heusler alloy Co$_2$MnSi is a \textit{full} Heusler alloy, i.e., it possesses two Co atoms per formula unit.
These Co atoms form cages in which eightfold coordinated Mn and Si atoms are enclosed [cf.~Fig.~\ref{fig:Heusler-Bands}(b)].
This is the so-called L$2_1$ structure; the corresponding space group is $Fm\bar{3}m$, which includes the inversion operation.
The experimental lattice constant is $5.654~\AA$~\cite{Webster:71}.
Here we will be using our theoretical USPP value, $a_0 = 5.633~\AA$, which is very close to the all-electron linearized augmented plane wave (LAPW) value $5.636~\AA$~\cite{Picozzi:02}.
Comparison of all-electron results displayed in Fig.~\ref{fig:Heusler-Bands}(c) (red diamonds) with our USPP band structure (black lines) further demonstrates the high quality of the pseudopotentials used here.

The most striking property of Co$_2$MnSi is its wide (indirect, $\Gamma$~--~$X$) band gap (energy width $\approx 0.81$~eV) in the minority spin channel,
which can clearly be seen in the band structure in Fig.~\ref{fig:Heusler-Bands}(c).
This special situation where one spin channel is metallic, while the other one is semiconducting or insulating, is referred to as \enquote{half-metallicity}~\cite{deGroot:83}.
The band gap in the minority spin channel is delimited by Co~$3d$ states
belonging to different representations of the symmetry group~\cite{Galanakis:02}.

Recent DFT calculations, complemented by the many-body quasiparticle $GW$~approximation, have corroborated the view of Co$_2$MnSi being a half-metallic ferromagnet~\cite{Meinert:12}.
A very recent experimental publication claims a large spin polarization of around $93\,\%$ in $70$~nm Co$_2$MnSi films grown epitaxially on MgO$(001)$ and a $30$~nm Co$_2$MnGa buffer layer on the basis of ultraviolet and x-ray photoemission spectroscopy experiments~\cite{Jourdan:14}.

A consequence of the half-metallicity is that, without inelastic processes, only majority spin electronic transport can occur around the Fermi energy.
Figure~\ref{fig:Heusler-FermiSurfaces} shows calculated Fermi surface sheets of the majority spin channel of bulk Co$_2$MnSi,
which can be helpful for the analysis of transport properties in the following.
According to Fig.~\ref{fig:Heusler-Bands}(c), the Fermi energy is crossed by three different bands along $\Gamma$~--~$X$ ($\Delta$ symmetry line).
However, while one band crosses the Fermi energy also along $\Gamma$~--~$L$ ($\Lambda$ symmetry line), which leads to a closed Fermi surface sheet, the other two bands cross the Fermi energy along $W$~--~$L$, which leads to Fermi surface sheets with \enquote{necks} along the $\langle 111 \rangle$~directions.

\subsection{Magnetic tunnel junctions}

\begin{figure}[b]
	\centering
	\includegraphics[]{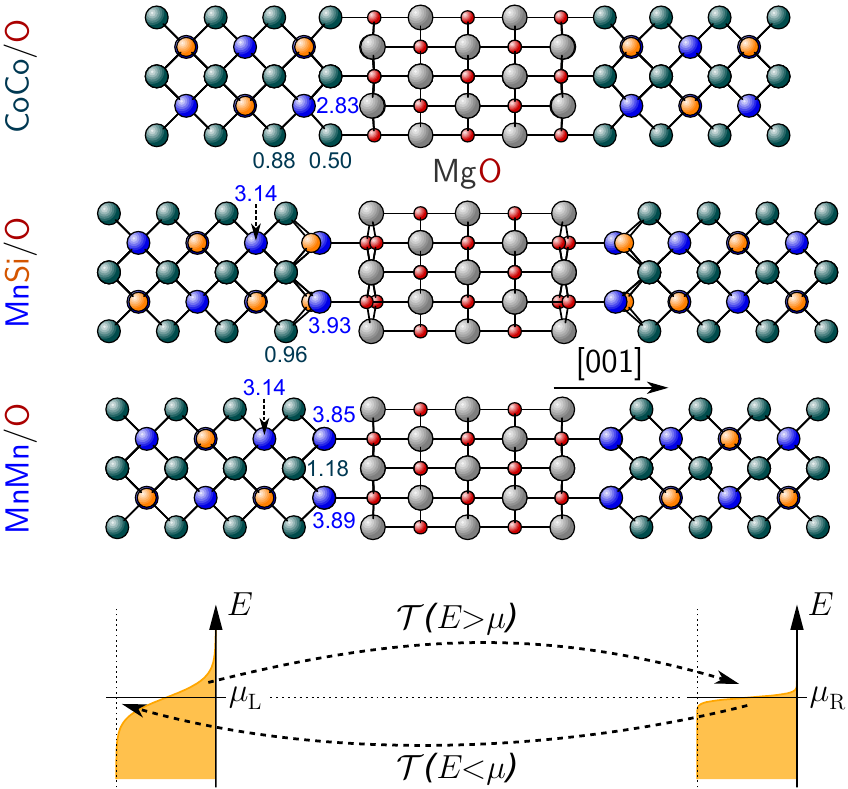}
	\caption{(Color online) Optimized atomic structure of epitaxial Co$_2$MnSi/MgO$(001)$/Co$_2$MnSi MTJs (shown with $5$ layers of MgO) with the three different interface terminations considered here: CoCo/O top, MnSi/O top, and MnMn/O top. The small numbers depict the local magnetic moments of Co and Mn atoms near the interface (in $\mu_{\text{B}}$). The bulk values for Co and Mn are $0.97$ and $3.17$~$\mu_{\text{B}}$, respectively. The lower image illustrates different thermal electron distributions in the two electrodes ($T_{\text{L}} > T_{\text{R}}$) and why a net current can flow in this case without an applied electric field, given that the transmission~ ${\cal T}(E)$ varies with the energy.}
	\label{fig:Heusler-TMR-Structures}
\end{figure}

The $(001)$ surface of Co$_2$MnSi can be matched epitaxially to MgO$(001)$ if either of them is rotated by $45^\circ$ about the $[001]$~axis.
The laterally smallest supercell that can be used to model the MTJs is tetragonal, with an in-plane lattice constant of $a_0 / \sqrt{2}$.
Thus, it contains Co$_2$MnSi in its rotated form.
Ideally, i.e., without the relaxation effects that occur in the vicinity of the interfaces, the tetragonal Heusler/MgO/Heusler supercells are set up such as to contain two atoms in each Heusler layer and four atoms in each MgO layer.

The calculated (experimental) lattice mismatch between Co$_2$MnSi and MgO is $6.6\,\%$ ($5.1\,\%$),
which means that the epitaxial MgO layer is subject to in-plane compressive strain.
Consequently, the MgO layer will distort tetragonally and expand in $[001]$~direction,
such that the distance between two atomic MgO$(001)$ layers increases by $5.6\,\%$ to $2.252\ \AA$.
The direct PBE band gap increases from $4.4$~eV to $5.0$~eV.

In Fig.~\ref{fig:Heusler-TMR-Structures} we can see the optimized atomic structure
of the epitaxial Co$_2$MnSi/MgO$(001)$ interface for the three different terminations considered here:
CoCo/O top, where both Co atoms sit on top of the O atoms of the insulator;
MnSi/O top, where the Mn and Si atoms sit on top of the O atoms;
and analogously MnMn/O top, which is a nonstoichiometric interface with Mn$_{\text{Si}}$ substitutions.
The selection of these three interfaces (out of many more different interface structures) is motivated by recent studies of the thermodynamic properties of different Co$_2$MnSi$(001)$ surfaces~\cite{Hashemifar:05} and different epitaxial Co$_2$MnSi/MgO$(001)$ interfaces~\cite{HuelsenPRL:09}.
In these works, it has been found that the CoCo/O and MnSi/O interfaces are the most stable ones, while the MnMn/O interface can be grown under nonequilibrium conditions and preserves the half-metallicity.

We find the CoCo/O and MnMn/O interfaces to be planar (cf.~Fig.~\ref{fig:Heusler-TMR-Structures}).
The Co-O bond length (given for $5$/$7$ layers of MgO) is the shortest with $2.10$/$2.10~\AA$ ($2.09~\AA$~\cite{HuelsenPRL:09}), while the Mn-O bond length is $2.38$/$2.35~\AA$ ($2.40~\AA$~\cite{HuelsenPRL:09}).
In contrast, the MnSi/O interface is corrugated: There seems to be some repulsion between Si and O atoms, which could be caused by a rehybridization of the under-coordinated interface Mn atoms.
Consequently, the Si atoms move towards the next CoCo layer in the Heusler electrode, while the Mn atoms form bonds with the O atoms.
The Si-O distance is $3.13$/$3.13~\AA$ ($3.17~\AA$~\cite{Miura:08}), and the Mn-O distance is $2.20$/$2.20~\AA$ ($2.27~\AA$~\cite{HuelsenPRL:09}, $2.25~\AA$~\cite{Miura:08}) now.
A similar effect occurs in the first MgO layer at the interface.
The electrode-electrode distance across the insulating spacer layer is largest for MnSi/O and smallest for CoCo/O (cf.~Fig.~\ref{fig:Heusler-TMR-Structures}).

Our interface band structures (not shown here) agree with previous all-electron results of H\"ulsen \textit{et al.}~\cite{HuelsenPRL:09}:
While the structurally optimized CoCo/O and MnSi/O interfaces induce interface states at the Fermi energy, the MnMn/O interface remains half-metallic.

Figure~\ref{fig:Heusler-TMR-Structures} does  not only show the atomic, but also the magnetic structure at the different interfaces.
The Co and Mn magnetic moments near a CoCo/O interface are significantly lowered when compared with bulk Co$_2$MnSi.
In contrast, the Mn moments are strongly increased close to a MnSi/O or MnMn/O interface.
The subsurface Co atoms exhibit bulklike (MnSi/O) or increased values (MnMn/O).

In total, there is a good agreement of the present structural, electronic, and magnetic results with those obtained by H\"ulsen \textit{et al.}\ (LAPW, Ref.~\cite{HuelsenPRL:09}) and Miura \textit{et al.}\ (USPP, Ref.~\cite{Miura:08}).
Small differences with respect to the literature arise mostly due to the different MgO barrier sizes used in the calculations ($3$~layers in Ref.~\cite{HuelsenPRL:09}, $5$/$7$~layers here), as we verified by performing calculations with different MgO barrier thickness.
Deviations from all-electron results due to the pseudopotential approach are found to be negligible.

\section{Electronic transport}

\begin{figure}[tbp]
	\centering
	\includegraphics[]{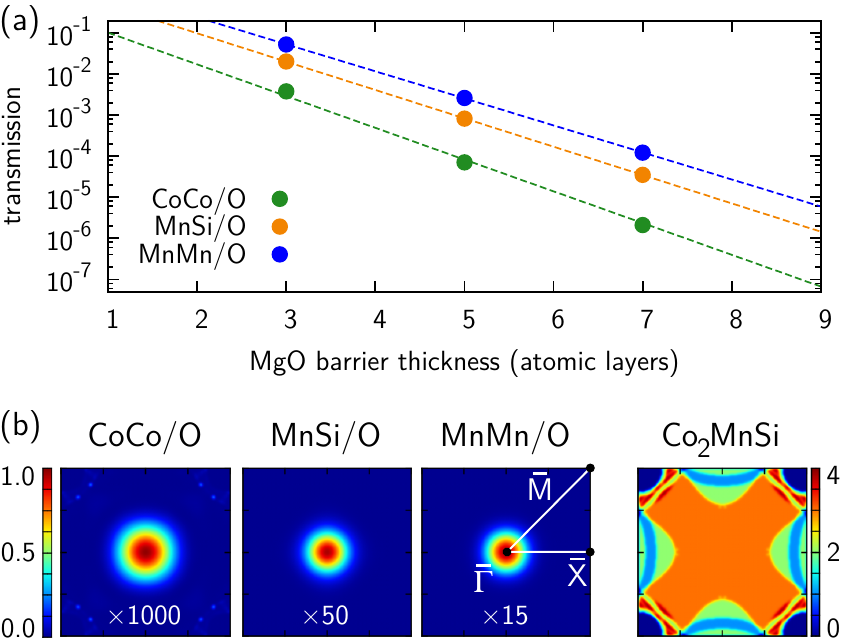}
	\caption{(Color online) (a)~Dependence of the majority spin transmission ${\cal T}_{\uparrow}(E_{\text{F}})$, Eq.~\eqref{eq:Heusler-TofE}, through Co$_2$MnSi/MgO$(001)$/Co$_2$MnSi MTJs on the MgO barrier thickness and on the interface termination. The exponential suppression of the transmission with increasing tunneling barrier thickness can be seen. (b)~Contour plots of the majority spin transmission ${\cal T}_{\uparrow}(E_{\text{F}}, \Vek{k}_{\perp})$ in the 2D BZ for $5$~layers of MgO and different interface terminations. Note the different scaling factors. The $\Vek{k}_{\perp}$-resolved transmission through a tetragonal unit cell of bulk Co$_2$MnSi is also shown. The 2D BZ refers to the tetragonal supercell, which contains the Heusler material in its $45^\circ$-rotated form.}
	\label{fig:Heusler-TMR-Transport-Dicke}
\end{figure}

\begin{figure}[tbp]
	\centering
	\includegraphics[]{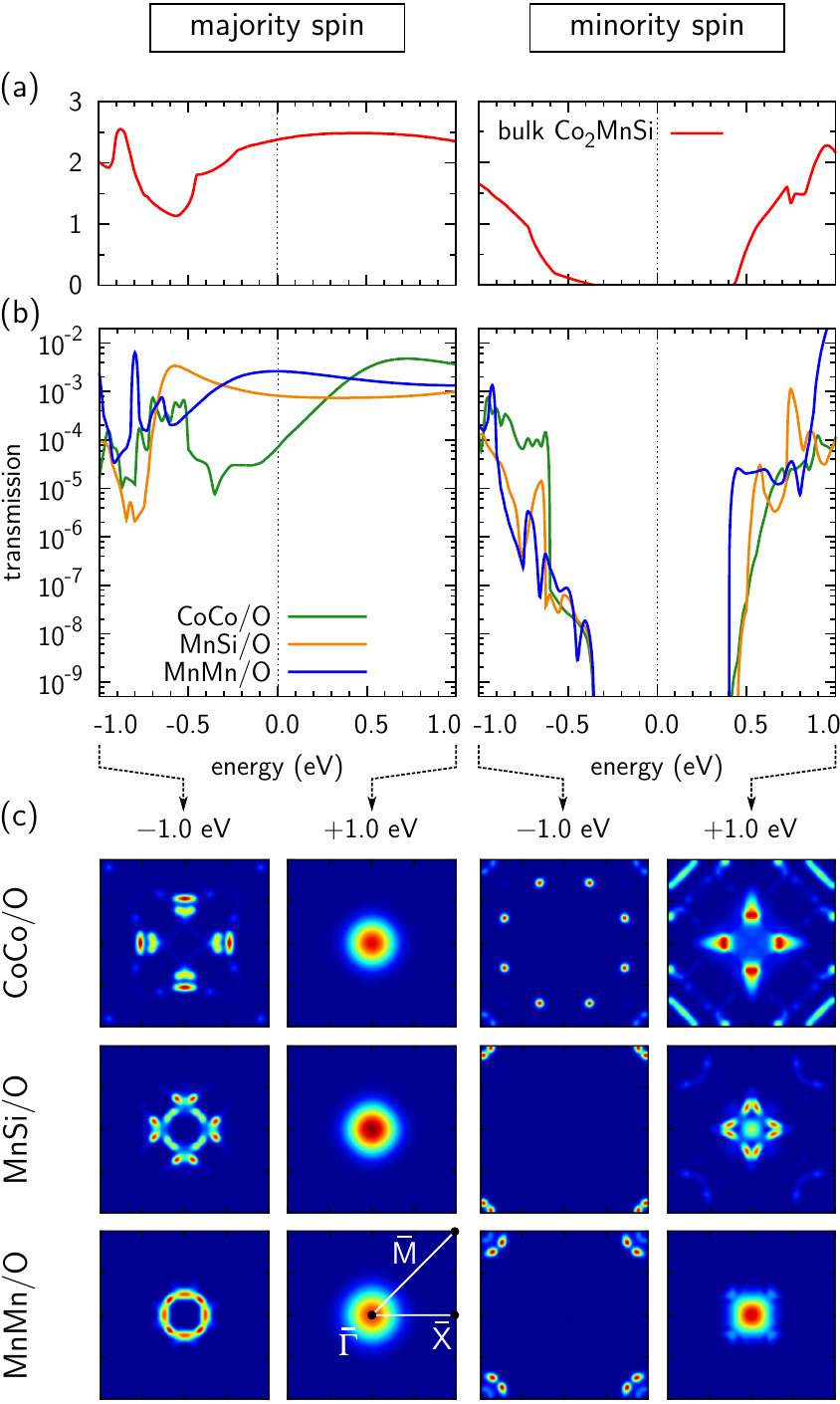}
	\caption{(Color online) Majority (left column) and minority (right column) spin transmission ${\cal T}_{\sigma}(E-E_{\text{F}})$ through (a)~bulk Co$_2$MnSi and (b)~Co$_2$MnSi/MgO$(001)$/Co$_2$MnSi MTJs with $5$~layers of MgO. (c)~Contour plots of ${\cal T}_{\sigma}(E, \Vek{k}_{\perp})$ for two selected energies relative to $E_{\text{F}}$. Note that all contour plots show tunneling transport and have different color scales (but fixed zero) to emphasize the \textit{qualitative} differences [cf.~Fig.~\ref{fig:Heusler-TMR-Transport-Dicke}(b)].}
	\label{fig:Heusler-TMR-TvonE}
\end{figure}

In the case of a parallel magnetization of the half-metallic electrodes, a conventional tunneling current can flow at least in the majority spin channel.
First, we will briefly comment on the transmission at the Fermi energy.
Afterwards, we will extend this view to a larger energy interval, as it is required for the subsequent determination of spincaloric properties.

In Fig.~\ref{fig:Heusler-TMR-Transport-Dicke}(a), we can see how the majority spin transmission at the Fermi energy ${\cal T}_{\uparrow}(E_{\text{F}})$, Eq.~\eqref{eq:Heusler-TofE}, depends on the MgO barrier thickness and on the interface termination.
For all three interfaces one observes an exponential decay of the transmission with the barrier thickness,
which is characteristic for tunneling through a potential barrier.
We use $3$ to $7$ atomic layers of MgO here.
Experimentally, barriers of $1.4$ to $3.2$~nm size have been used~\cite{Ishikawa:09, Liu:12}, corresponding roughly to $7$ to $15$ atomic layers of MgO.
This may lead to a further reduction of the transmission.
It is smallest for the CoCo/O interface and largest for the MnMn/O interface, which can be modeled by tunneling through potential barriers of different height (see below).
The contour plots of the majority spin transmission ${\cal T}_{\uparrow}(E_{\text{F}}, \Vek{k}_{\perp})$, which are displayed in Fig.~\ref{fig:Heusler-TMR-Transport-Dicke}(b) in the 2D BZ, show that the contributing channels are mostly concentrated around the $\bar{\Gamma}$~point (normal incidence).
However, there are also small satellite peaks in the vicinity of the $\bar{M}$~point in the case of the CoCo/O interface.
For comparison, the majority spin transmission through a tetragonal unit cell of bulk Co$_2$MnSi at the Fermi energy is also shown in Fig.~\ref{fig:Heusler-TMR-Transport-Dicke}(b).
It represents the available incoming or receiving transmission channels and is equivalent to the projected Fermi surface (cf.~Fig.~\ref{fig:Heusler-FermiSurfaces}). \footnote{The transmission has been obtained by considering an artificial system in which both the electrodes and the scattering region consist simply of bulk Co$_2$MnSi. It is an upper limit for the transmission through the MTJ cell. The same has been done to obtain the bulk transmission curves shown in Fig.~\ref{fig:Heusler-TMR-TvonE}(a).}

Aiming for spincaloric properties,
the transmission~${\cal T}_{\sigma}(E)$ needs to be calculated on a larger energy interval, not only at~$E_{\text{F}}$.
Since for each energy point the whole 2D BZ has to be sampled, this procedure is very time consuming.
The results in an energy interval of $\pm 1.0$~eV around $E_{\text{F}}$ are shown in Fig.~\ref{fig:Heusler-TMR-TvonE},
together with the energy-resolved transmission through a tetragonal unit cell of bulk Co$_2$MnSi, which can serve as reference.
Note that the whole energy range is within the MgO band gap (tunneling transport).
For the majority spin channel the transmission exhibits a smooth behavior for energies above $E_{\text{F}}-0.3$~eV regardless of the interface termination.
The appearance of several features below this energy coincides with the more structured majority spin band structure of bulk Co$_2$MnSi $0.5$~eV below its Fermi energy [cf.~Fig.~\ref{fig:Heusler-Bands}(c)].
Moreover, Fig.~\ref{fig:Heusler-TMR-TvonE}(b) clearly shows that the transmission curves cannot be matched by simply scaling the curves (i.e., exponentiation) or shifting the Fermi energies.
Since electrodes and barriers are equal in all systems, the \textit{qualitative} differences in the transmission properties can only stem from the interface termination.
This proves that the interface has a strong influence on the transport properties beyond simply modifying the MgO potential barrier height.
Moreover, we can see here that atomistic first-principles simulations
are a prerequisite for a more detailed understanding of quantum transport
that goes beyond conceptual studies.

The minority spin transmission vanishes around the Fermi energy due to the half-metallic band gap of the Heusler electrodes.
Beyond this gap, the minority spin transmission is mostly much smaller than the majority spin transmission.
Like the curves of the latter, those of the former exhibit a highly individual behavior, which points again to the influence of the interface.

The contour plots shown in Fig.~\ref{fig:Heusler-TMR-TvonE}(c) demonstrate that the structure of ${\cal T}_{\sigma}(E, \Vek{k}_{\perp})$ within the 2D BZ can be quite complex:
although the whole shown transmission through the barrier is due to tunneling,
it is in general \textit{not} concentrated around normal incidence ($\bar{\Gamma}$~point), in contrast to the findings at the Fermi energy.
Quite often there is no transmission at the $\bar{\Gamma}$~point at all.
Hence, it is not sufficient to restrict the 2D BZ sampling to the area around the $\bar{\Gamma}$~point (or even to use just the $\bar{\Gamma}$~point).
Consequently, the results of Miura \textit{et al.}~\cite{Miura:08} for the energy-resolved transmission ${\cal T}_{\sigma}(E)$ are quantitatively \textit{and} qualitatively different from our results,
even though their results for ${\cal T}_{\sigma}(E, \: \Vek{k}_{\perp}\!\!=\!\bar{\Gamma})$ agree with those obtained by us
under the same assumptions
for testing purposes (not shown here).

\begin{figure}[tbp]
	\centering
	\includegraphics[]{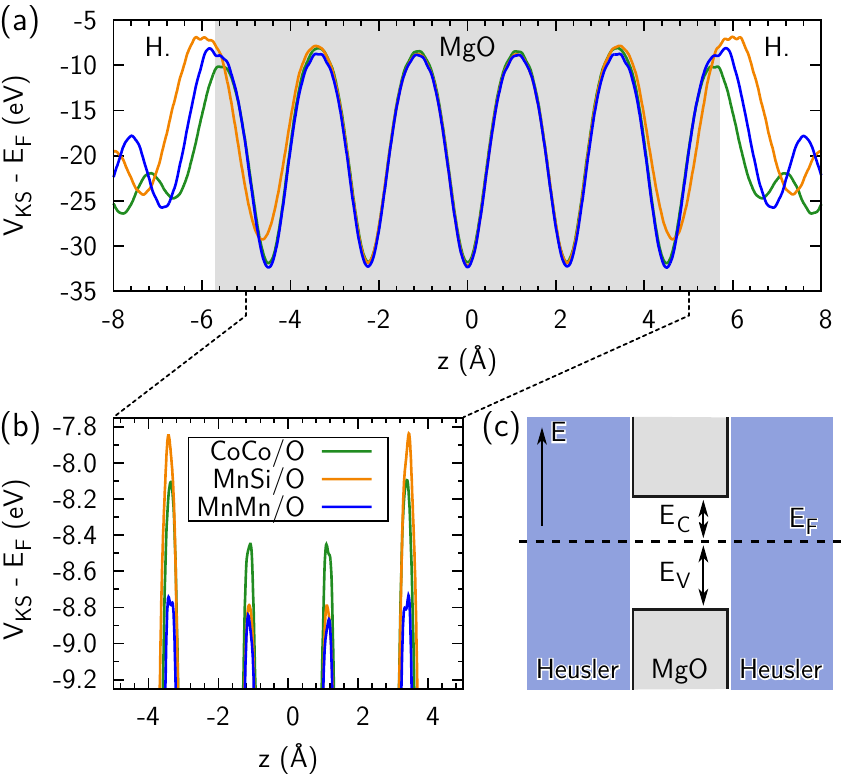}
	\caption{(Color online) (a)~Majority spin Kohn-Sham potentials $V_{\text{KS}}(\Vek{r})$ of MTJs with $5$~layers of MgO (gray shaded region) between Co$_2$MnSi Heusler electrodes (labeled by \enquote{H.}; only the interface region is shown). The potentials have been averaged in the $xy$~plane and aligned such that the Fermi energies $E_{\text{F}}$ of the corresponding systems coincide [$V_{\text{KS}}(\Vek{r})=0$~eV here]. (b)~Magnification of~(a). (c)~Scheme of the electronic structure around the Fermi energy to illustrate the definitions of $E_{\text{V}}$ and $E_{\text{C}}$.}
	\label{fig:Heusler-Potentials}
\end{figure}

It is not straightforward to explain the strong dependence of the energy-resolved transmission ${\cal T}_{\uparrow}(E)$ on the interface termination.
Figures~\ref{fig:Heusler-Potentials}(a) and~\ref{fig:Heusler-Potentials}(b) show the majority spin Kohn-Sham potentials $V_{\text{KS}}(\Vek{r})$, which have been averaged in the $xy$~plane and aligned such that the Fermi energies of the corresponding systems coincide, for the three different MTJs.
Incoming electrons have to traverse these potentials and are scattered differently.
Two aspects shall be discussed here:
(i)~On the one hand, the potential peak sequence in the \textit{central barrier region} is CoCo/O, MnSi/O, and MnMn/O,
which agrees with the reversed sequence
of the transmission magnitude observed at the Fermi energy [cf.~Fig.~\ref{fig:Heusler-TMR-Transport-Dicke}(a)].
One can extract the energy separations
between the MTJ Fermi energy and the MgO valence band maximum $E_{\text{V}}$
or the conduction band minimum $E_{\text{C}}$ [cf.~Fig.~\ref{fig:Heusler-Potentials}(c)]
by matching the potentials in the central barrier region
with the $xy$-averaged potential of tetragonally distorted bulk MgO.
The $E_{\text{C}}$ values for CoCo/O, MnSi/O, and MnMn/O are
$2.28$, $1.92$, and $1.82$~eV,
respectively,
whereas the corresponding $E_{\text{V}}$ values are
$2.72$, $3.08$, and $3.18$~eV.
This shows explicitly that the energy interval considered in Fig.~\ref{fig:Heusler-TMR-TvonE}
is indeed located \textit{within} the band gap of epitaxial MgO.
We find that the ratios of different $\sqrt{E_\text{C}}$ agree
with the ratios of the fitted slopes $\kappa$ in Fig.~\ref{fig:Heusler-TMR-Transport-Dicke}(a),
in the spirit that ${\cal T}(E_{\text{F}}) \sim \Exp^{-\kappa d}$, where $\kappa \sim \sqrt{E_{\text{C}}}$ and $d$ is the barrier thickness.
(ii)~On the other hand, the broadest and highest (smallest and lowest) potential peak at the \textit{interface} belongs to the MnSi/O (CoCo/O) termination [cf.~Fig.~\ref{fig:Heusler-Potentials}(a)], which also exhibits the largest (smallest) electrode-insulator spacing in its atomic structure (cf.~Fig.~\ref{fig:Heusler-TMR-Structures}).
This reflects the influence of the different bonding at the interface and the different atomic species involved.
The interplay of all these aspects determines the energy-resolved transmission curves.

\section{Spincaloric properties}

\subsection{Calculation of the Seebeck coefficients}

In the regime of linear response,
where temperature gradients and voltages are assumed to be infinitesimally small,
the total current can be expressed as
\begin{equation}
\label{eq:LinRespCurrent}
I = \left( \Delta \mu / e - S \Delta T \right) \Multp G
\text{,}
\end{equation}
where $\Delta \mu = \mu_{\text{L}} - \mu_{\text{R}}$ and $\Delta T = T_{\text{L}} - T_{\text{R}}$.
The conductance~$G$ and the Seebeck coefficient~$S$ arising in this equation can be obtained by using the approach of Sivan and Imry~\cite{SI86},
which starts from the central quantity ${\cal T}_\sigma(E)$
and the Fermi distribution function $f = f_{\mu, T}(E)$.
Within Mott's two-current model, the spin-projected and temperature-dependent conductance is expressed as
\begin{equation}
  \label{eq:Heusler-Conductance}
  G_\sigma(T)  = -\frac{e^2}{h} \,
        \int \text{d}E \, \frac{\partial f}{\partial E} \,
        {\cal T}_\sigma(E)
\text{,}
\end{equation}
the total conductance being simply $G = G_{\uparrow} + G_{\downarrow}$,
and the spin-projected Seebeck coefficients take on the form 
\begin{equation}
  \label{eq:Heusler-Seebeck}
  S_\sigma(T)  = -\frac{1}{eT} \,
        \frac{{\displaystyle \int \text{d}E \,
           \frac{\partial f}{\partial E} \,
               (E-\mu) \, {\cal T}_\sigma(E)}}
             {{\displaystyle\int \text{d}E \,
         \frac{\partial f}{\partial E} \, {\cal T}_\sigma(E)}}
\text{.}
\end{equation}
They are not additive ($S_{\uparrow} + S_{\downarrow} \neq S$) due to the different denominators
and do not have a strict physical meaning.
However, with these quantities
the effective (\enquote{charge}) and the spin-dependent Seebeck coefficient can be expressed as
\begin{equation}
  \label{eq:Heusler-EffSpinSeeb}
  S_{\text{eff}} = \frac{G_\uparrow\,S_\uparrow +
                         G_\downarrow\,S_\downarrow}
                    {G_\uparrow + G_\downarrow}
\quad
\text{and}
\quad
  S_{\text{spin}} = \frac{G_\uparrow\,S_\uparrow -
                         G_\downarrow\,S_\downarrow}
                    {G_\uparrow + G_\downarrow}
\text{.}
\end{equation}
Thus, the two spin channels are treated as parallel connected resistors here.
The spin-dependent Seebeck coefficient is a measure for the thermally induced spin accumulation.
In Eqs.~\eqref{eq:Heusler-Conductance} and~\eqref{eq:Heusler-Seebeck} one usually sets $\mu \equiv E_{\text{F}}$, where $E_{\text{F}}$ denotes the common Fermi energy of the MTJ cell and of the electrodes, thereby neglecting any temperature dependence of the chemical potential.
The precise meaning of the temperature~$T$, commonly regarded as average temperature~\cite{SI86}, will become transparent later,
since actually there should be \textit{two} temperatures, $T_{\text{L}}$ and~$T_{\text{R}}$, corresponding to the left and the right lead, respectively.

\begin{figure}[tbp]
	\centering
	\includegraphics[]{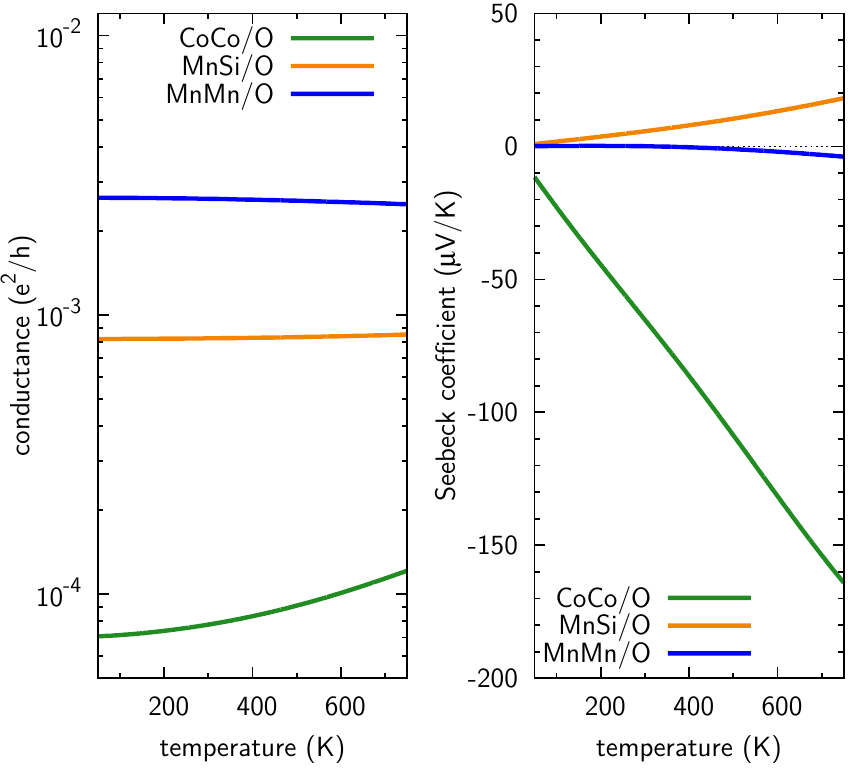}
	\caption{(Color online) Results obtained with the Sivan-Imry approach for Co$_2$MnSi/MgO$(001)$/Co$_2$MnSi MTJs with different interface terminations and $5$~layers of MgO. The left panel shows the conductance $G_{\uparrow}(T) \approx G(T)$ [Eq.~\eqref{eq:Heusler-Conductance}]. The right panel shows the Seebeck coefficient $S_{\text{eff}}(T) \approx S_{\uparrow}(T) \approx S_{\text{spin}}(T)$ [Eqs.~\eqref{eq:Heusler-Seebeck} and~\eqref{eq:Heusler-EffSpinSeeb}].}
	\label{fig:Heusler-TMR-G-S}
\end{figure}

Results for the epitaxial Co$_2$MnSi/MgO$(001)$/Co$_2$MnSi MTJs are shown in Fig.~\ref{fig:Heusler-TMR-G-S}.
For the temperature range considered here there are no relevant contributions to~$G$ from the minority spin channel ($G_{\downarrow}/G_{\uparrow} < 2.5 \Multp 10^{-4}$) due to the half-metallic band gap in the electrodes.
Thus, $S_{\text{eff}} \approx S_{\uparrow} \approx S_{\text{spin}}$,
which means that the entire voltage generated under a temperature gradient is converted into a spin accumulation.
The conductance is largest for the MnMn/O interface and smallest for the CoCo/O interface.
In contrast, the CoCo/O interface leads to the largest Seebeck coefficient (in absolute value), while it is smallest for the MnMn/O interface.

These results, especially the latter one, can be anticipated from the transmission curves shown in Fig.~\ref{fig:Heusler-TMR-TvonE}(b).
Equation~\eqref{eq:Heusler-Seebeck} makes it clear that the Seebeck coefficient strongly depends on the asymmetry of the transmission~${\cal T}_\sigma(E)$ around the Fermi energy~$E_{\text{F}}$.
If, for instance, the Fermi energy lies in a band gap, the Seebeck coefficient can be tailored by shifting the Fermi energy towards one of the band edges.
This can be done by doping or adequate selection of the electrode materials.
Another (equivalent) route is a shifting of the entire band gap (valence band maximum and conduction band minimum) around the Fermi energy, which can be done, for instance, by using different Heusler spacer layers between a fixed electrode material, or by exploiting the band structure modifications induced by epitaxial strain~\cite{GeislerPopescu:14, ComtesseGeisler:14}.

\begin{table}[b]
	\centering
	\caption{\label{tab:Heusler-SeebeckComparison}Comparison of the effective and spin-dependent Seebeck coefficients determined for the present system to results for other Co$_2$-Heusler-based systems with a CoCo interface structure around $T=300$~K from the literature.}
	\begin{ruledtabular}
	\begin{tabular}{lcc}
		System, Interface & $S_{\text{eff}}$ ($\mu$V/K) & $S_{\text{spin}}$ ($\mu$V/K)	\\
		\hline
		Co$_2$MnSi/MgO/Co$_2$MnSi, CoCo/O				&  $-65$				& $-65$		\\
		Al/Co$_2$Ti(Si,Ge)/Al, CoCo~\cite{GeislerPopescu:14}		&  $-13$ to $+1$		& $-3$ to $+1$		\\
		Pt/Co$_2$(Mn,Fe)(Si,Al)/Pt, CoCo~\cite{ComtesseGeisler:14}	&  $-4$ to $+18$		& $-5$ to $+5$		\\
	\end{tabular}
	\end{ruledtabular}
\end{table}

Comparison of the results we obtained for our Co$_2$MnSi/MgO$(001)$/Co$_2$MnSi MTJs to other Co$_2$-Heusler-based systems with CoCo interfaces that have been investigated recently reveals that the Seebeck coefficients for the present system are much higher (cf.~Table~\ref{tab:Heusler-SeebeckComparison}).
For the CoCo/O interface, our Seebeck coefficients are even higher than those for Fe/MgO/Fe or Co/MgO/Co MTJs~\cite{Heiliger-FeMgO-Spincalorics:11}.
This leads to larger and more easily detectable thermally induced voltages, as we will also see in more detail below, which makes the considered system more attractive for applications.
Comparison with the recently measured, small Seebeck coefficient of bulk Co$_2$MnSi ($-6$~$\mu$V/K at $300$~K~\cite{Balke:10})
underlines the advantage of such a nanostructured MTJ.

\begin{figure}[tbp]
	\centering
	\includegraphics[]{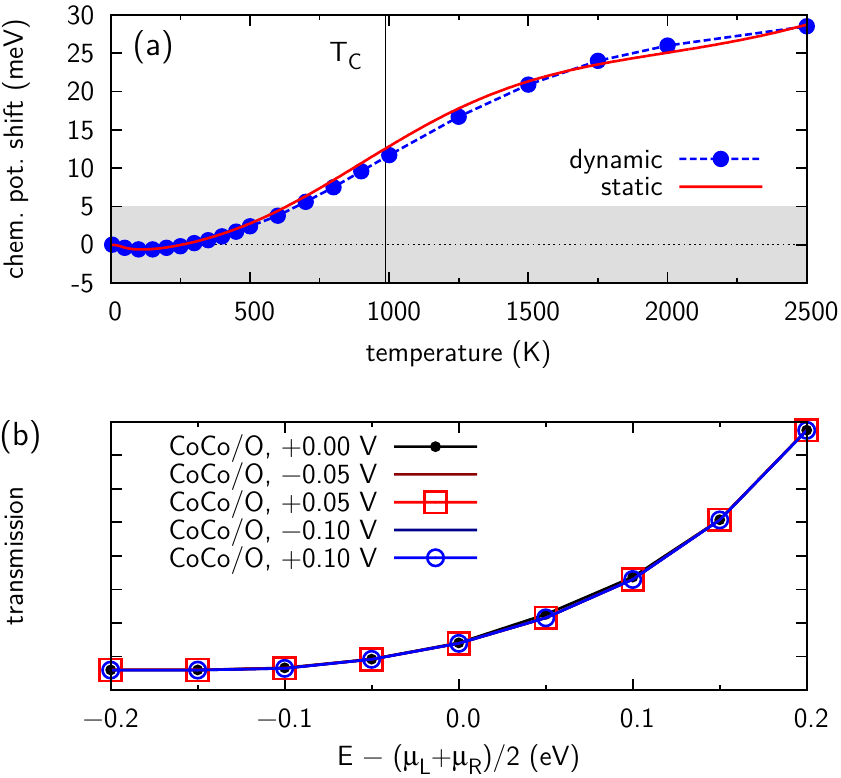}
	\caption{(Color online) (a)~Variation of the chemical potential~$\mu(T)$ with the temperature in bulk Co$_2$MnSi, calculated in two different ways as explained in the text (\enquote{static} vs.~\enquote{dynamic}). (b)~Finite-bias influence of small voltages $V \in [-0.1,+0.1]$~V on the transmission~${\cal T}(E,V)$ through a Co$_2$MnSi/MgO$(001)$/Co$_2$MnSi MTJ with CoCo/O interfaces and $5$~layers of MgO. One can see that effects of finite bias voltages are negligible here.}
	\label{fig:Heusler-TMR-Fermi-FiniteBias}
\end{figure}

It is worthwhile to speculate
what will happen if the MgO band gap is increased, e.g., due to a more accurate \textit{ab initio} description of its electronic structure.
A larger $E_{\text{C}}$ value [cf.~Fig.~\ref{fig:Heusler-Potentials}(c)] will suppress the transmission, $\tilde{\cal T}_{\sigma}(E) = \alpha(E) {\cal T}_{\sigma}(E)$, which lowers the conductance.
However, the suppression $\alpha$ will not be a constant factor [which would cancel in the calculation of the Seebeck coefficient, Eq.~\eqref{eq:Heusler-Seebeck}], but a function of the energy, and, as an\-tici\-pated from the analytical expression for tunneling through a rectangular potential barrier, stronger for higher than for lower energies.
The thereby induced change of the asymmetry around the Fermi energy will scale the Seebeck coefficients towards less negative / more positive values.
For the CoCo/O interface, for instance, we estimated a change from $-65$ to $-60~\mu$V/K at $300$~K\footnote{We assume $\tilde{E}_{\text{C}} = E_{\text{C}}+1.5$~eV due to a symmetrically increased MgO band gap of experimental width (nearly $8$~eV~\cite{Madelung:04}).
Analytical expressions for tunneling through a rectangular potential barrier of $12~\AA$ thickness
imply $\alpha(E)\approx\frac{1}{100} \left[ 1-\frac{9}{10}\frac{E-E_{\text{F}}+1~\text{eV}}{2~\text{eV}} \right]$ in the energy region studied here [Fig.~\ref{fig:Heusler-TMR-TvonE}(b)].
We rescaled our transmission curves accordingly and deduced the \enquote{corrected} Seebeck coefficient.}.

There are two drawbacks to the procedure used so far.
First, the temperature dependence of the chemical potentials in the electrodes
has not been accounted for.
Indeed, this aspect is frequently neglected in \textit{ab initio} studies~\cite{Mavropoulos:15, Mavropoulos:14, GeislerPopescu:14, ComtesseGeisler:14, Popescu:13, Heiliger-FeMgO-Spincalorics:11}.
It is known that for semiconductors, for instance, this temperature dependence
is crucial for the thermoelectric properties.
We calculated it (i)~by populating a fixed zero-temperature band structure according to the Fermi-Dirac distribution for different temperatures (called \enquote{static} here)
and, for comparison, (ii)~during the DFT self-consistent field runs by using a Fermi-Dirac smearing with different temperatures (called \enquote{dynamic} here).
Both methods lead to quite similar results for Co$_2$MnSi [cf.~Fig.~\ref{fig:Heusler-TMR-Fermi-FiniteBias}(a)],
which means that the response of the electronic system to the temperature-increased smearing is small.
For the temperatures of interest here, the shift of the chemical potential~$\mu(T)$ is smaller than~$\pm 5$~meV.
It is tempting to use this as justification to neglect it
during the calculation of the system's response to a temperature gradient;
whether this is acceptable or not will be discussed in the following in the context of a different approach.

As a second disadvantage of the linearized treatment,
finite-bias effects due to the potential difference between the two electrodes cannot be included.
We investigated this aspect for the present MTJs and found that
the influence of small, but finite voltages (and thermally induced voltages are usually small)
between the electrodes is negligible
[cf.~Fig.~\ref{fig:Heusler-TMR-Fermi-FiniteBias}(b)],
which is probably related to the uniform behavior of the bands near $E_{\text{F}}$ [cf.~Fig.~\ref{fig:Heusler-Bands}(c)].
The shown transmission curves~${\cal T}(E,V)$ have been calculated for several bias voltages~$V$ from symmetrically shifted bands in the electrodes around the common Fermi energy.
Due to the symmetric setup of the MTJ cell, the sign of the bias voltage has no influence on the transmission.
If finite-bias effects are not negligible,
the approach outlined in the following is capable of including them.

\subsection{An alternative route to spincaloric properties}

The procedure used so far is currently the standard route to calculating thermoelectric and/or spincaloric properties.
For instance, it has been used recently to investigate Al/Co$_2$TiSi/Al and Al/Co$_2$TiGe/Al heterostructures~\cite{GeislerPopescu:14}.
However, this formalism is only an approximation and works best for \textit{very small} thermal gradients between the two contacts.
It is, though, possible to access thermoelectric and/or spincaloric properties more exactly
and \textit{without} calculating a Seebeck coefficient at all;
this is presented in the following.

\begin{figure}[tbp]
	\centering
	\includegraphics[]{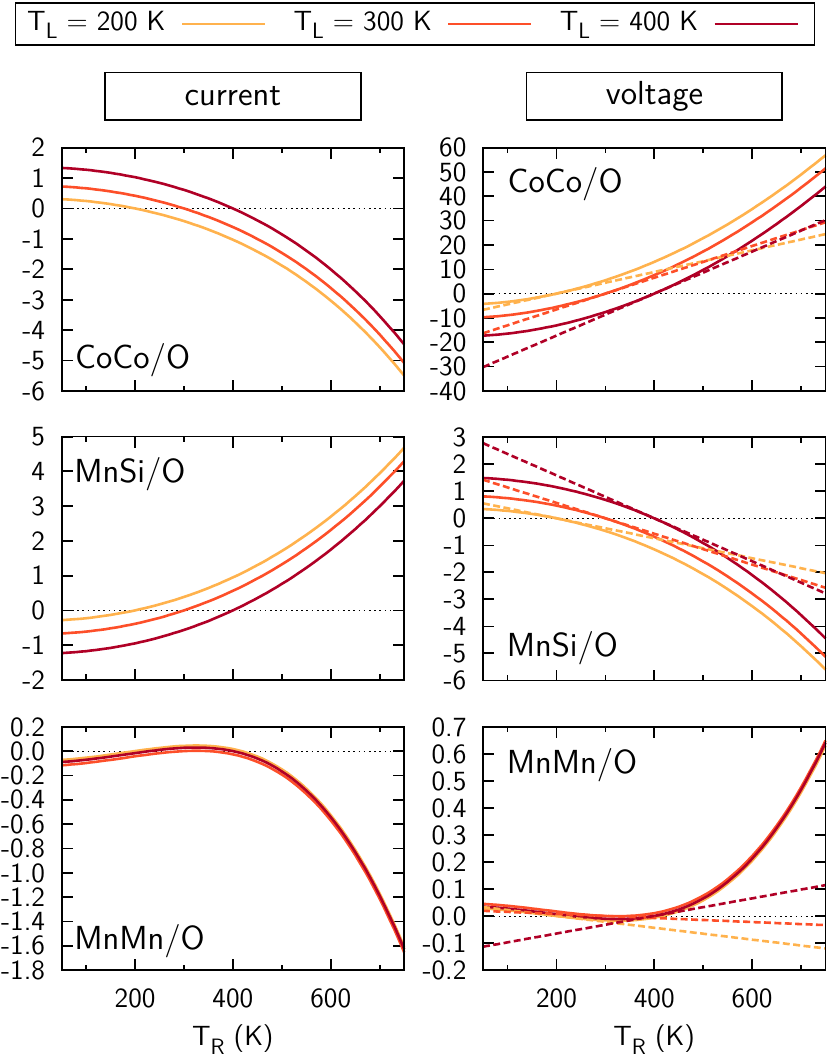}
	\caption{(Color online) (Left) Thermally driven closed-circuit currents~$I(T_{\text{L}},T_{\text{R}})$ ($e^2/h \Multp \mu$V) as calculated from Eq.~\eqref{eq:Heusler-LBI} for three Co$_2$MnSi/MgO$(001)$/Co$_2$MnSi MTJs with different interfaces and $5$~layers of MgO. (Right) Comparison of the open-circuit voltages~$V(T_{\text{L}},T_{\text{R}}) = (\mu_{\text{L}} - \mu_{\text{R}})/e$ (mV) as determined from solving Eq.~\eqref{eq:Heusler-LBV} (solid lines) and corresponding voltages calculated from $S_{\text{eff}}(T_{\text{L}})$ by using Eq.~\eqref{eq:Heusler-SeebeckTaylor} (dashed lines) for the same three systems.}
	\label{fig:Heusler-TMR-V-I}
\end{figure}

In the end one is interested in a current~$I$ or a voltage~$V = (\mu_{\text{L}} - \mu_{\text{R}})/e$ arising as a response of the MTJ to an applied thermal gradient or, more precisely, to the two applied temperatures $T_{\text{L}}$ and $T_{\text{R}}$ in the left and the right electrode, respectively.
If the circuit is closed, a thermally driven current~$I$ will flow, which can be calculated directly from the Landauer-B\"uttiker formula:
\begin{equation}
\label{eq:Heusler-LBI}
I(T_{\text{L}},T_{\text{R}}) = \frac{e}{h} \Int{}{E} \left[ f_{T_{\text{L}}}(E) - f_{T_{\text{R}}}(E) \right] {\cal T}(E)
\text{,}
\end{equation}
where ${\cal T} = {\cal T}_{\uparrow} + {\cal T}_{\downarrow}$.
Since no counteracting electric field can build up ($V=0$), it follows that $\mu_{\text{L}} = \mu_{\text{R}} = E_{\text{F}}$;
thus, the chemical potentials have been omitted in the formula.
The currents calculated for our Co$_2$MnSi-based MTJs with $5$~layers of MgO can be seen in Fig.~\ref{fig:Heusler-TMR-V-I}, left column.

If we consider, on the other hand, an open circuit without a current, $I=0$,
the charge flow induced by the thermal gradient has to be compensated by an electric field, which is proportional to~$V$.
By using the Landauer-Büttiker formula once more, we can now write:
\begin{equation}
\label{eq:Heusler-LBV}
0 \overset{!}{=} \frac{e}{h} \Int{}{E} \left[ f_{\mu_{\text{L}}, T_{\text{L}}}(E) - f_{\mu_{\text{R}}, T_{\text{R}}}(E) \right] {\cal T}(E)
\text{.}
\end{equation}
The goal is to find a pair $(\mu_{\text{L}}, \, \mu_{\text{R}})$ that solves this integral equation, which parametrically depends on the temperatures $T_{\text{L}}$ and $T_{\text{R}}$.
Since the potential drop across the devices studied here will be symmetric due to their symmetric construction, the following additional assumption is reasonable:
\begin{equation*}
(\mu_{\text{L}} + \mu_{\text{R}})/2 = E_{\text{F}}
\text{,}
\end{equation*}
which can be used to eliminate one of the variable chemical potentials.
Besides, this reduces the numerical effort required to solve Eq.~\eqref{eq:Heusler-LBV}.

We note that the transmission~${\cal T}(E)$ in Eq.~\eqref{eq:Heusler-LBV} is not recalculated for appropriately shifted bands here (finite-bias effects).
Therefore, generally speaking, even this approach is an approximation which works best for small response voltages~$V$.
Luckily, thermally induced voltages are usually small enough, especially in the present temperature range.
Moreover, it is shown explicitly in Fig.~\ref{fig:Heusler-TMR-Fermi-FiniteBias}(b) that finite-bias effects can safely be neglected here.
If, in contrast, the differences between ${\cal T}(E)$ and ${\cal T}(E,V)$ were not negligible,
such finite-bias effects could be included in the present approach.
An improved version of Eq.~\eqref{eq:Heusler-LBV} would read:
\begin{equation*}
0 \overset{!}{=} \frac{e}{h} \Int{}{E} \left[ f_{\mu_{\text{L}}, T_{\text{L}}}(E) - f_{\mu_{\text{R}}, T_{\text{R}}}(E) \right] {\cal T}(E, \frac{\mu_{\text{L}} - \mu_{\text{R}}}{e})
\text{,}
\end{equation*}
but is computationally even more demanding than the aforementioned approach, since ${\cal T}(E,V)$ has to be calculated for several bias voltages~$V$.\footnote{In Ref.~\onlinecite{Heiliger-FeMgOFe:06}, for instance, which addresses \textit{field-driven} transport through Fe/MgO/Fe MTJs, ${\cal T}(E,V)$ was calculated for several bias voltages~$V$ from symmetrically shifted bands.}
Especially in conjunction with thermally induced transport, which requires larger energy integration intervals than field-driven transport, this can be tedious.

\begin{figure}[tbp]
	\centering
	\includegraphics[]{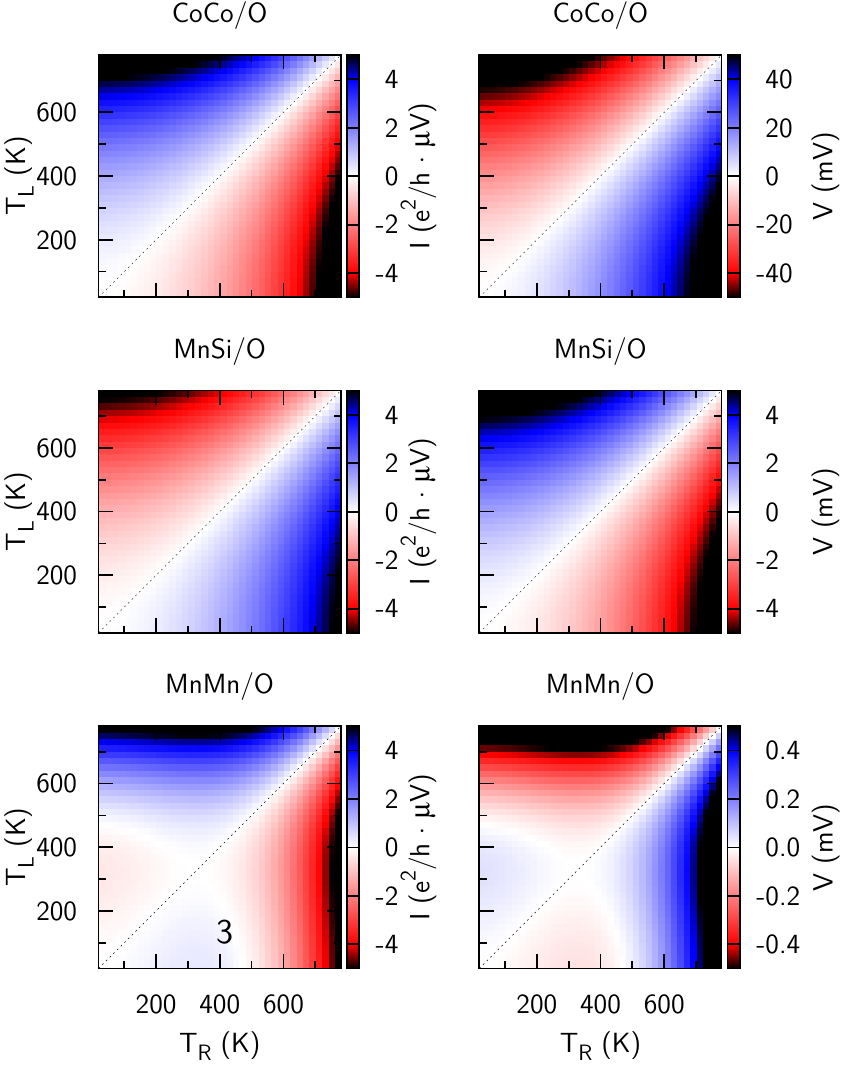}
	\caption{(Color online) Thermally driven closed-circuit currents~$I(T_{\text{L}},T_{\text{R}})$ as calculated from Eq.~\eqref{eq:Heusler-LBI} (left) and open-circuit voltages~$V(T_{\text{L}},T_{\text{R}}) = (\mu_{\text{L}} - \mu_{\text{R}})/e$ as determined from solving Eq.~\eqref{eq:Heusler-LBV} (right) for Co$_2$MnSi/MgO$(001)$/Co$_2$MnSi MTJs with different interfaces and $5$~layers of MgO.}
	\label{fig:Heusler-TMR-V-I-Panels}
\end{figure}

The solution of Eq.~\eqref{eq:Heusler-LBV} can be found, for instance, by using numerical integration and the bisection method starting from $\mu_{\text{L}} = \mu_{\text{R}}  = E_{\text{F}}$.
This provides the voltage response~$V(T_{\text{L}},T_{\text{R}})$ of the MTJ as shown in Fig.~\ref{fig:Heusler-TMR-V-I}, right column.
On the other hand, the Sivan-Imry Seebeck coefficient~$S_{\text{eff}}(T)$ defined in Eq.~\eqref{eq:Heusler-EffSpinSeeb} can be understood as first-order Taylor expansion coefficient of this voltage,
\begin{equation}
\label{eq:Heusler-SeebeckTaylor}
V(T_{\text{L}}^{},T_{\text{R}}^{}) = S_{\text{eff}}^{}(T_{\text{L}}^{}) \Multp (T_{\text{L}}^{} - T_{\text{R}}^{}) + \mathcal{O}(T_{\text{R}}^{2})
\text{.}
\end{equation}
This becomes obvious from Fig.~\ref{fig:Heusler-TMR-V-I}, right column, where the dashed lines are tangent to the real voltage curves around~$T_{\text{R}} = T_{\text{L}}$.
One can see explicitly here that the Sivan-Imry approach provides quite good results for small thermal gradients, as expected.

Both current~$I$ and voltage~$V$ are given in Fig.~\ref{fig:Heusler-TMR-V-I-Panels} for several combinations of $T_{\text{L}}$ and $T_{\text{R}}$.
Due to the symmetric setup of the MTJ cell, all panels are antisymmetric with respect to the dashed diagonal line ($T_{\text{L}} = T_{\text{R}}$).
The thermally induced voltages do not exceed $70$~mV for the considered temperature range,
which justifies the neglect of finite-bias effects \textit{a posteriori}.
There is no simple linear dependence between $I$ and $V$, although one could get this impression from Fig.~\ref{fig:Heusler-TMR-V-I-Panels}.
The generated voltage~$V$ is largest for the CoCo/O interface and smallest for the MnMn/O interface.
Moreover, the sign of $I$ and $V$ is reversed for MnSi/O with respect to CoCo/O and MnMn/O, which is also consistent with the conventionally determined Seebeck coefficients shown in Fig.~\ref{fig:Heusler-TMR-G-S},
as is the sign flip that can be observed in current and voltage for the MnMn/O interface.

\subsection{The role of the electrode chemical potentials}

So far, we have neglected the temperature dependence of the chemical potentials in the electrodes,
$\mu_{\text{L}}(T_{\text{L}})$ and $\mu_{\text{R}}(T_{\text{R}})$.
It has been shown in Fig.~\ref{fig:Heusler-TMR-Fermi-FiniteBias}(a)
that the variation of the chemical potential with the temperature is quite small for Co$_2$MnSi.
On the other hand,
we can expect an approximate voltage correction
$\Delta \mu / e = \mu_{\text{L}}(T_{\text{L}})/e - \mu_{\text{R}}(T_{\text{R}})/e$
in the spirit of Eq.~\eqref{eq:LinRespCurrent},
which can be of similar size as the voltages calculated
for MnSi/O and MnMn/O (cf.~Figs.~\ref{fig:Heusler-TMR-V-I} and~\ref{fig:Heusler-TMR-V-I-Panels}).
In the following, we will use our approach to obtain the exact correction. 

Now, the finite temperature does not only broaden the reservoirs' Fermi distribution functions,
but also shifts them slightly to higher or lower energies.
The current can be expressed as
\begin{equation}
\label{eq:Heusler-LBI-MU}
\tilde{I}(T_{\text{L}},T_{\text{R}}) = \frac{e}{h} \Int{}{E} \left[ f_{\mu_{\text{L}}(T_{\text{L}}), T_{\text{L}}}(E) - f_{\mu_{\text{R}}(T_{\text{R}}), T_{\text{R}}}(E) \right] {\cal T}(E)
\text{,}
\end{equation}
very similar to Eq.~\eqref{eq:Heusler-LBI}.
In analogy to Eq.~\eqref{eq:Heusler-LBV}, the integral equation from which the voltage can be calculated reads:
\begin{equation}
\label{eq:Heusler-LBV-MU}
0 \overset{!}{=} \frac{e}{h} \Int{}{E} \left[ f_{\mu_{\text{L}}(T_{\text{L}})+\lambda, T_{\text{L}}}(E) - f_{\mu_{\text{R}}(T_{\text{R}})-\lambda, T_{\text{R}}}(E) \right] {\cal T}(E)
\text{.}
\end{equation}
In this context, the $\mu_{\text{L/R}}(T)$ are not variables, but predetermined functions
providing solely the thermal variation of the electrode chemical potentials,
whereas $\lambda$ models the field-induced potential shift and is determined by using the bisection method, very similar to the case investigated above.
From this quantity, the voltage follows as $\tilde{V}(T_{\text{L}},T_{\text{R}}) = 2 \lambda / e$.
The formalism supports electrodes made from different materials,
but since both electrodes are made from the same material here, we can set $\mu_{\text{L}}(T) = \mu_{\text{R}}(T) = \mu(T)$.
We use the $\mu(T)$ in the following that has been obtained with the \enquote{static} method [cf.~Fig.~\ref{fig:Heusler-TMR-Fermi-FiniteBias}(a)].

\begin{figure}[tbp]
	\centering
	\includegraphics[]{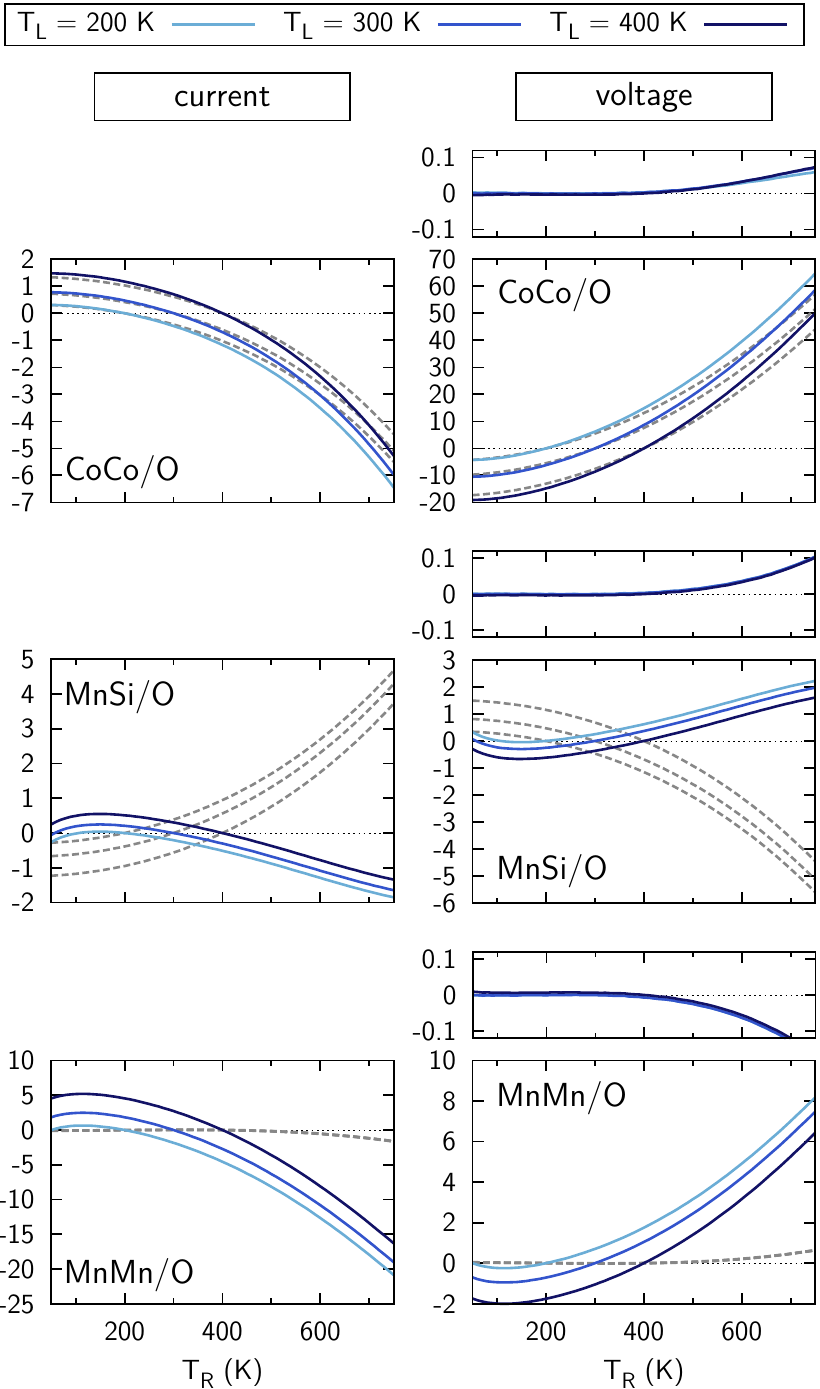}
	\caption{(Color online) Influence of the temperature-dependent chemical potentials in the leads. (Left) Thermally driven closed-circuit currents~$\tilde{I}(T_{\text{L}},T_{\text{R}})$ ($e^2/h \Multp \mu$V) as calculated from Eq.~\eqref{eq:Heusler-LBI-MU} for three Co$_2$MnSi/MgO$(001)$/Co$_2$MnSi MTJs with different interfaces and $5$~layers of MgO. (Right) The open-circuit voltages~$\tilde{V}(T_{\text{L}},T_{\text{R}})$ (mV) as determined from solving Eq.~\eqref{eq:Heusler-LBV-MU} (solid lines) for the same three systems. The gray dashed lines in each panel are replicated current and voltage curves from Fig.~\ref{fig:Heusler-TMR-V-I} for comparison. The small panels show the difference $\tilde{V}(T_{\text{L}},T_{\text{R}}) - V(T_{\text{L}},T_{\text{R}}) - \left( \mu(T_{\text{R}}) - \mu(T_{\text{L}}) \right) / e$ (mV) for each of the three curves.}
	\label{fig:Heusler-TMR-V-I-MU}
\end{figure}

\begin{figure}[tbp]
	\centering
	\includegraphics[]{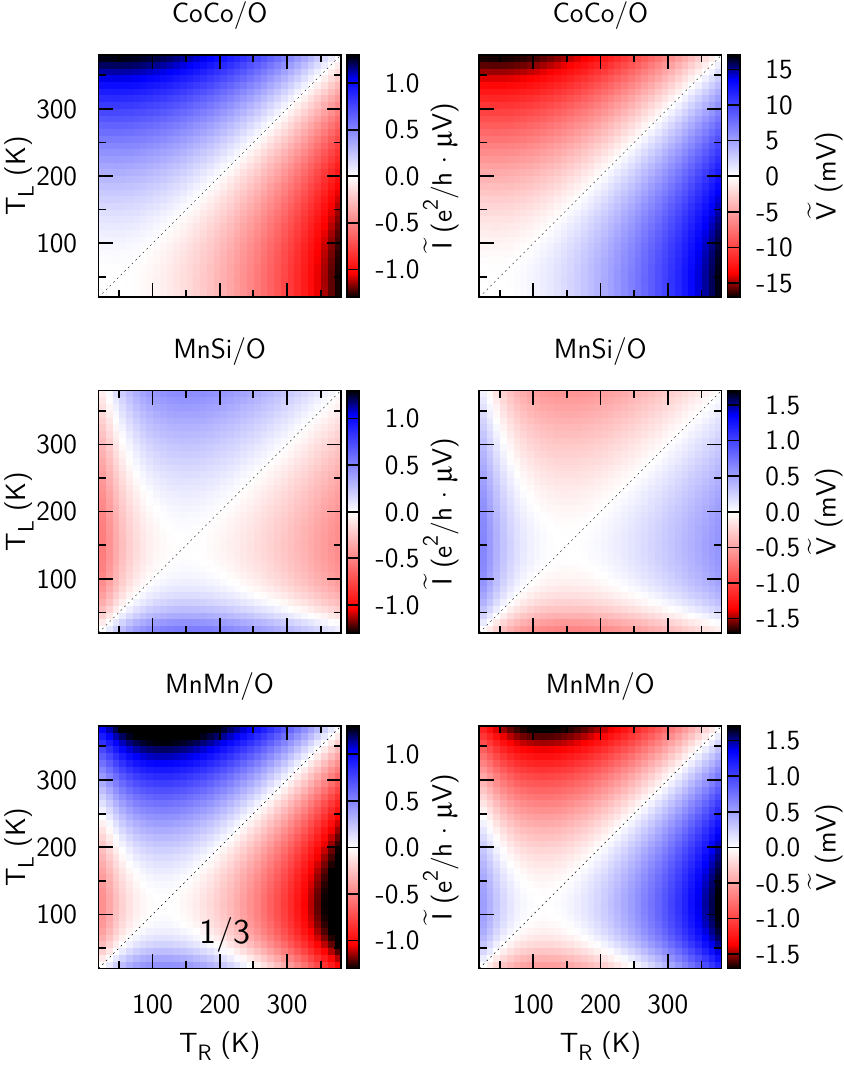}
	\caption{(Color online) Thermally driven closed-circuit currents~$\tilde{I}(T_{\text{L}},T_{\text{R}})$ as calculated from Eq.~\eqref{eq:Heusler-LBI-MU} (left) and open-circuit voltages~$\tilde{V}(T_{\text{L}},T_{\text{R}})$ as determined from solving Eq.~\eqref{eq:Heusler-LBV-MU} (right) for Co$_2$MnSi/MgO$(001)$/Co$_2$MnSi MTJs with different interfaces and $5$~layers of MgO, including the effects of temperature-dependent chemical potentials in the electrodes. The temperature range is smaller than in Fig.~\ref{fig:Heusler-TMR-V-I-Panels} to underline the differences.}
	\label{fig:Heusler-TMR-V-I-MU-Panels}
\end{figure}

As we can see in Fig.~\ref{fig:Heusler-TMR-V-I-MU},
the influence of the temperature-dependent chemical potentials on current and voltage is very strong,
even though $\mu(T)$ varies only in the small range of $\pm 5$~meV.
While the relative corrections to current and voltage are small for the CoCo/O interface,
sign and magnitude are changed for the MnSi/O and, especially, the MnMn/O interface.
This can also be seen in Fig.~\ref{fig:Heusler-TMR-V-I-MU-Panels} for several combinations of $T_{\text{L}}$ and $T_{\text{R}}$.
We end up with changes of quantitative \textit{and} qualitative nature.
The differences $\tilde{I}-I$ and $\tilde{V}-V$ are found to correspond to the largest part
to $\Delta \mu / e \Multp G$ and $-\Delta \mu / e$, respectively,
a behavior similar to linear response [cf.~Eq.~\eqref{eq:LinRespCurrent}].
This is illustrated in the small panels in Fig.~\ref{fig:Heusler-TMR-V-I-MU}.
While Eqs.~\eqref{eq:Heusler-LBI-MU} and~\eqref{eq:Heusler-LBV-MU} in fact do provide further corrections,
as they are caused by the interplay of thermal shift and broadening of the Fermi distribution functions
in conjunction with the nonconstant transmission,
these nontrivial contributions are found to be of minor importance here.
For the curves shown in Fig.~\ref{fig:Heusler-TMR-V-I-MU}, they are close to zero for $T_{\text{R}} < 300$~K;
we find that they play a role only for higher temperatures, i.e., for a larger
broadening of the expression $f_{\mu_{\text{L}}, T_{\text{L}}} - f_{\mu_{\text{R}}, T_{\text{R}}}$.

In order to get a mathematical impression of these corrections,
we first approximate the transmission around the Fermi energy (chosen as zero) roughly as
\begin{equation*}
{\cal T}(E) = \tau_0 + \tau_1 E + \mathcal{O}(E^{2})
\text{,}
\end{equation*}
where $\tau_0 = {\cal T}(0)$ and $\tau_1 = {\cal T}'(0)$,
and subsequently perform a Sommerfeld expansion of Eq.~\eqref{eq:Heusler-LBV-MU}:
%
\begin{eqnarray*}
0 &\overset{!}{=}& \int_{\mu_{\text{R}}-\lambda}^{\mu_{\text{L}}+\lambda} \hspace{-2em} \text{d}E \, {\cal T}(E)
+ \frac{\pi^2}{6} k_{\text{B}}^2
\left[ T_{\text{L}}^2 {\cal T}'(\mu_{\text{L}}\!+\!\lambda) - T_{\text{R}}^2 {\cal T}'(\mu_{\text{R}}\!-\!\lambda) \right]	\\
&=& \left( 2\lambda \!+\! \mu_{\text{L}} \!-\! \mu_{\text{R}} \right) \left[ \tau_0 + \frac{\mu_{\text{L}} \!+\! \mu_{\text{R}}}{2} \tau_1 \right]
+ \frac{\pi^2}{6} k_{\text{B}}^2
\tau_1 \left( T_{\text{L}}^2 \!-\! T_{\text{R}}^2 \right)
\text{,}
\end{eqnarray*}
%
where $\mu_{\text{L}} \equiv \mu(T_{\text{L}})$ and $\mu_{\text{R}} \equiv \mu(T_{\text{R}})$.
This equation can be solved algebraically for $2 \lambda \approx e \tilde{V}(T_{\text{L}},T_{\text{R}})$.
Hence, the difference between the voltages obtained from Eqs.~\eqref{eq:Heusler-LBV-MU} and~\eqref{eq:Heusler-LBV} for the linear model transmission is:
%
\begin{equation*}
\tilde{V} - V \approx - \frac{\Delta \mu}{e}
- \frac{\pi^2}{6} \frac{k_{\text{B}}^2}{e} \tau_1 \left( T_{\text{L}}^2 \!-\! T_{\text{R}}^2 \right)
\left\lbrace \frac{1}{\tau_0 \!+\! \frac{\mu_{\text{L}}+\mu_{\text{R}}}{2} \tau_1} \!-\! \frac{1}{\tau_0} \right\rbrace
\text{.}
\end{equation*}
%
For finite temperature gradients, the second term vanishes only if $\tau_1 = 0$ or $\mu(T) \equiv 0$
and therefore always provides a correction to the first term, even for this simple model transmission.

We conclude that accounting for the temperature-dependent chemical potentials in the electrodes
is crucial in order to get both the correct current and voltage response of the system,
since the thermally induced voltages are of the same order of magnitude
as the variations of the chemical potentials.
The conventional Seebeck coefficient~$S_{\text{eff}}$ alone and the voltage it implies
according to Eq.~\eqref{eq:Heusler-SeebeckTaylor} can be misleading.
Note that the variations of $\mu(T)$ in popular electrode materials like Fe or Al are of the same order of magnitude
as they are in Co$_2$MnSi.

\subsection{Thermally operated MRAM modules}

\begin{table}[t]
	\centering
	\caption{\label{tab:Heusler-TMRAM-Voltages}Exemplary voltages generated by single MTJ cells (parallel electrode magnetization) with different interfaces, operated at different temperatures $T_{\text{L}}$/$T_{\text{R}}$. The values in parentheses take the temperature-dependent chemical potentials into account [Eq.~\eqref{eq:Heusler-LBV-MU}].}
	\begin{ruledtabular}
	\begin{tabular}{lcc}
		Interface & $290$/$310$~K & $340$/$360$~K	\\
		\hline
		CoCo/O	&  $+1.30$~mV ($+1.48$~mV)		& $+1.51$~mV ($+1.72$~mV)		\\
		MnSi/O	&  $-0.11$~mV ($+0.06$~mV)		& $-0.14$~mV ($+0.07$~mV)		\\
		MnMn/O	&  $-1.48~\mu$V ($+0.18$~mV)	& $+1.84~\mu$V ($+0.21$~mV)		\\
	\end{tabular}
	\end{ruledtabular}
\end{table}

\begin{figure}[b]
	\centering
	\includegraphics[]{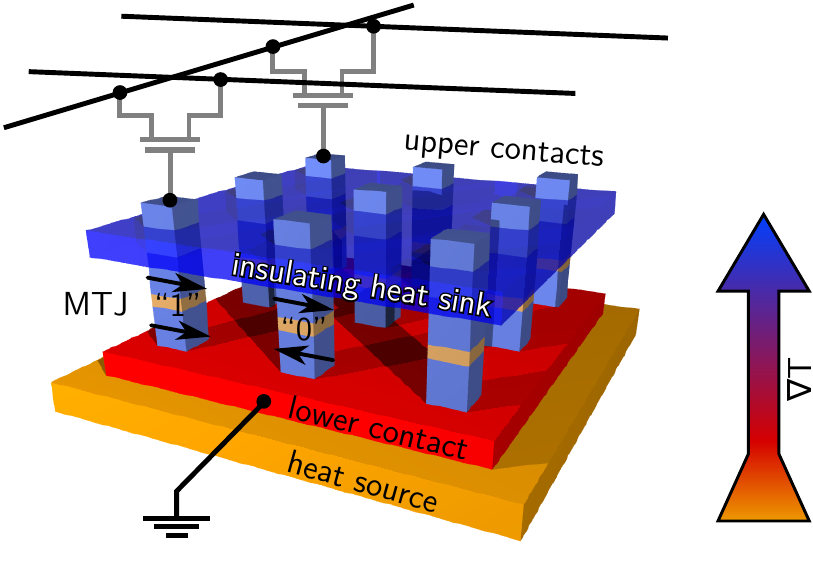}
	\caption{(Color online) Illustration of the suggested thermo-MRAM module in the cross-point architecture based on the magneto-Seebeck effect. The thermal gradient between the heat source and the heat sink generates different voltages in the MTJs which depend on their magnetic state (parallel/antiparallel electrode magnetization) and can be used to detect the state of a selected MTJ and thus the information stored in it. Writing units are not shown here.}
	\label{fig:Heusler-TMR-thermo-MRAM}
\end{figure}

We end this paper with a practical example.
Table~\ref{tab:Heusler-TMRAM-Voltages} shows some voltages generated by a single MTJ if the two electrodes (parallel magnetization) are operated at different temperatures, for instance, around room temperature.
We see here explicitly that
the magnitude (and, at lower operating temperatures, also the sign; cf.~Fig.~\ref{fig:Heusler-TMR-V-I-MU-Panels})
of the thermally induced voltage can be tailored
by exploiting the fact that the MTJ interface formation can be controlled by adjusting the growth conditions~\cite{HuelsenPRL:09}.
Since especially the CoCo/O voltages can be measured without problems, Co$_2$MnSi/MgO$(001)$/Co$_2$MnSi MTJs,
grown under Co-rich conditions,
can be used in future \enquote{thermo-MRAM} modules (cf.~Fig.~\ref{fig:Heusler-TMR-thermo-MRAM}),
where the stored information is read out without a flowing charge current
by exploiting the magneto-Seebeck effect.
This is different in conventional MRAM modules~\cite{Akerman:05}.
In the case of a parallel electrode magnetization (state~\enquote{1}),
a voltage will be generated by the MTJ
(due to the temperature gradient between the heat source and the heat sink)
that acts upon the gate of a field-effect transistor.
If the electrodes are magnetized antiparallel (state~\enquote{0}), no (or, at least, a much lower) voltage arises,
and the transistor remains blocked.
The application of a thermal gradient is only necessary for the readout process;
the stored information is not lost if $T_{\text{L}} = T_{\text{R}}$. 
Moreover, it can be exploited that
the current (and thus the possible power) scales with the area of the MTJ,
whereas the voltage can be increased, if necessary, by a serial arrangement of MTJs.
We think that such devices could be used in modern, energy-efficient computers,
where, for example, the heat emitted by the CPU, in conjunction with its cooling heat sink, provides the necessary temperature gradient.

\section{Summary}

We have investigated
the electronic transport and spincaloric properties of
epitaxial magnetic tunnel junctions with half-metallic Co$_2$MnSi Heusler electrodes, MgO tunneling barriers, and different interface terminations on the basis of first-principles calculations.
It has been shown that the interface has a strong influence on the electronic transport properties
beyond simply modifying the height of the tunneling barrier potential,
and that the tunneling transmission is not necessarily concentrated around the $\bar{\Gamma}$~point in the two-dimensional Brillouin zone.
We have calculated the transmission on a large energy interval for each interface,
and from these results the spincaloric properties have been obtained with the linearized method of Sivan and Imry.
For comparison,
a new approach has been presented that circumvents the linear response approximation inherent in the Seebeck coefficient.
This approach supports two temperatures with finite difference in the two electrodes
and provides the exact current and/or voltage response of the system.
Moreover, it can directly account for temperature-dependent chemical potentials in the electrodes and finite-bias effects,
and we have shown that especially the former are important for obtaining qualitatively correct results,
even if the variations of the chemical potentials are small in the present system.
It has been suggested how the spincaloric properties can be tailored by the choice of the growth conditions.
In particular, we have found a large effective and spin-dependent Seebeck coefficient of $-65$~$\mu$V/K at room temperature
for the purely Co-terminated interface.
Such interfaces can be used in thermally operated magnetoresistive random access memory modules,
which are based on the magneto-Seebeck effect and which we have suggested here, to maximize the thermally induced readout voltage.

\begin{acknowledgments}

This work was supported by the German Research Foundation
(Deutsche Forschungsgemeinschaft, DFG) within the
Priority Program 1538 \enquote{Spin Caloric Transport (SpinCaT).}

\end{acknowledgments}

\end{document}